\documentclass[twocolumn,aps,superscriptaddress,floatfix]{revtex4-1}
\usepackage{array}
\usepackage{amsmath}
\usepackage{amssymb}
\usepackage{graphicx}
\usepackage{color}
\usepackage{adjustbox}
\usepackage{lipsum}

\newcommand{\be}{\begin{equation}}
\newcommand{\ee}{\end{equation}}
\newcommand{\bea}{\begin{eqnarray}}
\newcommand{\eea}{\end{eqnarray}}
\newcommand{\bse}{\begin{subequations}}
\newcommand{\ese}{\end{subequations}}

\newcommand{\ems}{EuMg$_2$Sb$_2$}
\newcommand{\emb}{EuMg$_2$Bi$_2$}
\newcommand{\esa}{EuSn$_2$As$_2$}
\newcommand{\nsa}{NaSn$_2$As$_2$}
\newcommand{\cas}{CaAl$_2$Si$_2$}

\begin{document}

\title{Antiferromagnetic domains in a single crystal of the A-type spin-7/2 trigonal topological insulator EuSn$_2$As$_2$}

\author{Santanu Pakhira}
\affiliation{Ames National Laboratory, Iowa State University, Ames, Iowa 50011, USA}
\affiliation{Present Address: Institute for Quantum Materials and Technologies,  Karlsruhe Institute of Technology, 76131, Karlsruhe, Germany.}
\author{D. C. Johnston}
\affiliation{Ames National Laboratory, Iowa State University, Ames, Iowa 50011, USA}
\affiliation{Department of Physics and Astronomy, Iowa State University, Ames, Iowa 50011, USA}

\date{\today}

\begin{abstract}

EuSn$_2$As$_2$ is a trigonal A-type antiferromagnetic topological insulator with the moments aligned in the $ab$ plane and with a N\'eel temperature $T_{\rm N} = 23.5$~K\@.  Here we report that an EuSn$_2$As$_2$ crystal exhibits a broad peak at $H_{\rm c1} = 1100$~Oe in the field derivative $dM_{ab}/dH$of the $ab$-plane magnetization $M_{ab}(H)$  at temperature $T=2$~K, demonstrating the presence of trigonal antiferromagnetic domains.  We model these $M_{ab}(H,\,T=2\,{\rm K})$ data and obtain the trigonal anisotropy-energy coefficient $K_3$ that is 13.3 and 3.7 times larger than those we previously reported for single crystals of the trigonal compounds \ems\ and \emb, respectively.

\end{abstract}

\maketitle

\section{Introduction}

\begin{figure}[ht]
\includegraphics[width = 3.in]{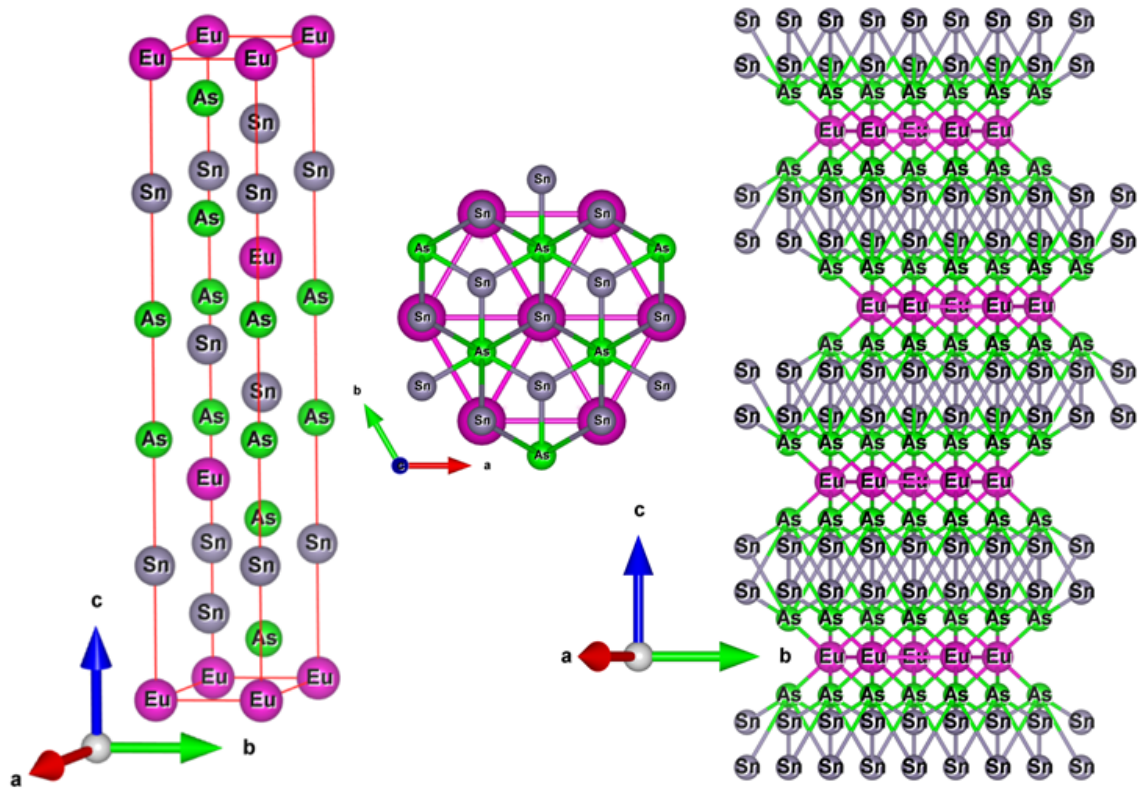}
\caption{Trigonal crystal structure of \esa~\cite{Pakhira2021}.  The Eu atoms are colored dark purple, the Sn atoms blue, and the As atoms green.  The Eu atoms form a hexagonal (triangular) layer situated between two buckled SnAs honeycomb layers, with four Eu layers within each $c$-axis height.  The Eu magnetic structure is A-type with the Eu$^{2+}$ moments aligned in the $ab$~plane \cite{Pakhira2021}.  The height of the magnetic unit cell is twice the $c$-axis lattice parameter.}
\label{Fig_Xtal_Struct}
\end{figure}

\begin{figure}[ht]
\includegraphics[width = 2.5in]{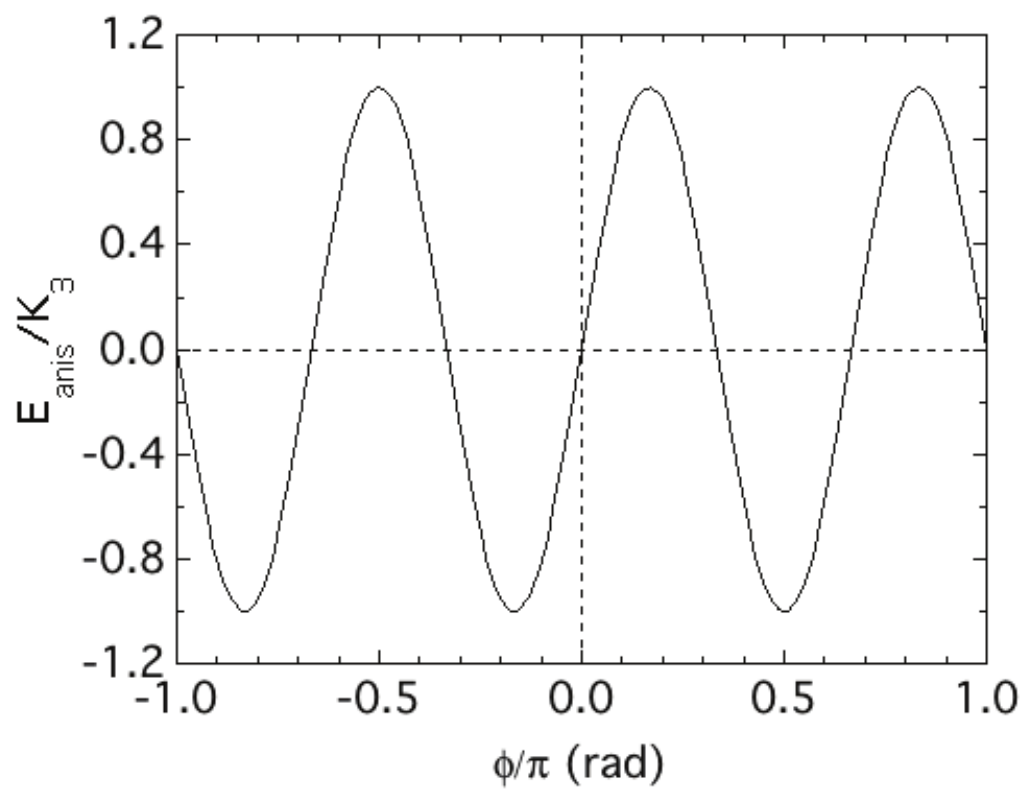}
\caption{Magnetic anisotropy free energy $E_{\rm anis}$ of ferromagnetic moments aligned at an angle $\phi$ with the positive $x$~axis, normalized by the trigonal anisotropy constant $K_3$ in Eq.~(\ref{Eq:FofPhi}).}
\label{FigTrigAnisEnergy}
\end{figure}

\esa\ is a topological insulator that has been reported to exhibit Dirac surface states at energies within the bulk band gap from angle-resolved photoemission spectroscopy (ARPES) measurements~\cite{Li2019_EuSn2As2}.  The compound crystallizes in the trigonal \nsa-type structure with space group $R\overline{3}m$ shown in Fig.~\ref{Fig_Xtal_Struct}~\cite{Pakhira2021}.  It orders antiferromagnetically below its N\'eel temperature $T_{\rm N} \approx 23.5$~K~\cite{Pakhira2021, Arguilla2017_EuSn2As2}.  Neutron-diffraction measurements~\cite{Pakhira2021} showed that the antiferromagnetic (AFM) structure is \mbox{A-type,} where the Eu$^{+2}$ moments have spin $S=7/2$, spectroscopic splitting factor $g=2$, and ordered moment $\mu = gS\mu_{\rm B}=7\mu_{\rm B}$, where $\mu_{\rm B}$ is the Bohr magneton.  The Eu moments are aligned ferromagnetically in an $ab$~plane layer with the Eu moments in adjacent layers along the $c$~axis aligned antiferromagnetically~\cite{Pakhira2021}.  Other studies of the magnetic and thermodynamic properties of  \esa\ single crystals were also presented in Ref.~\cite{Pakhira2021}.

\begin{figure}[ht]
\includegraphics[width = 1.8in]{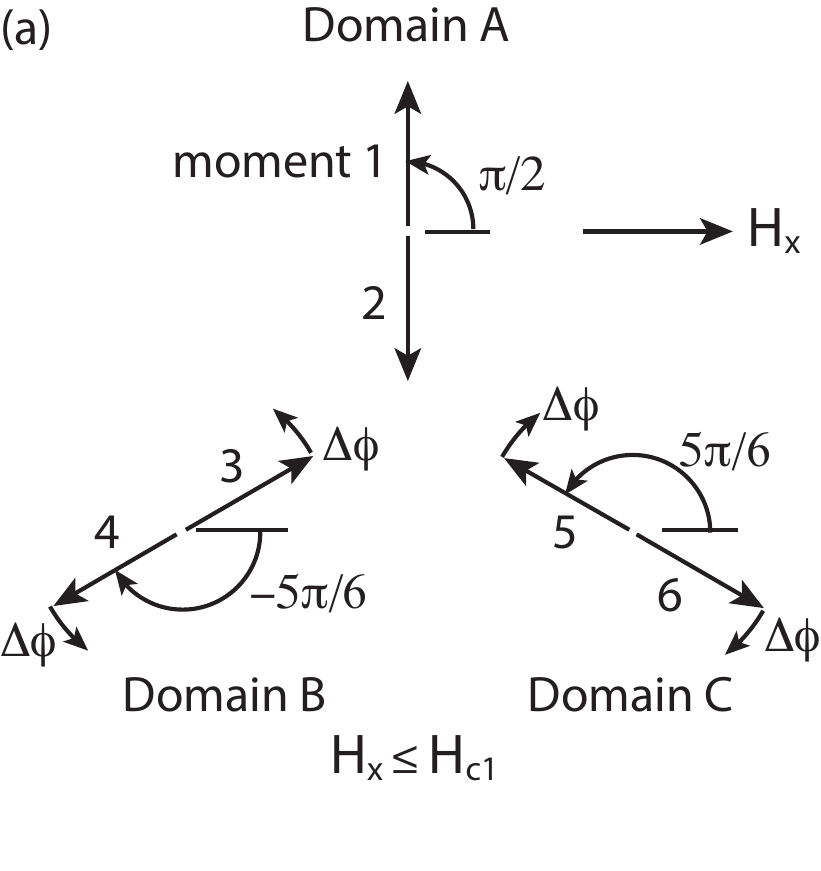}
\includegraphics[width = 1.8in]{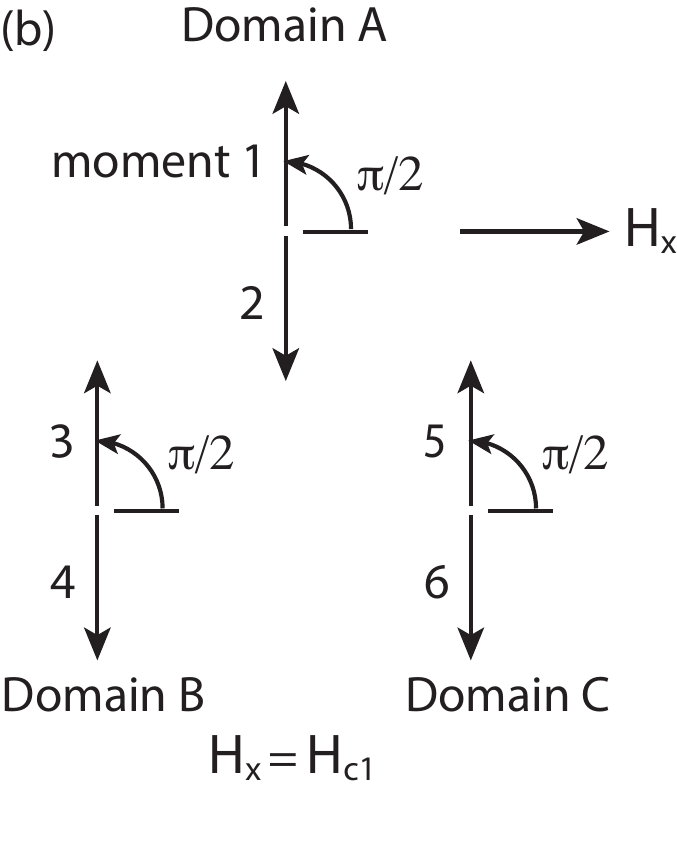}
\includegraphics[width = 1.8in]{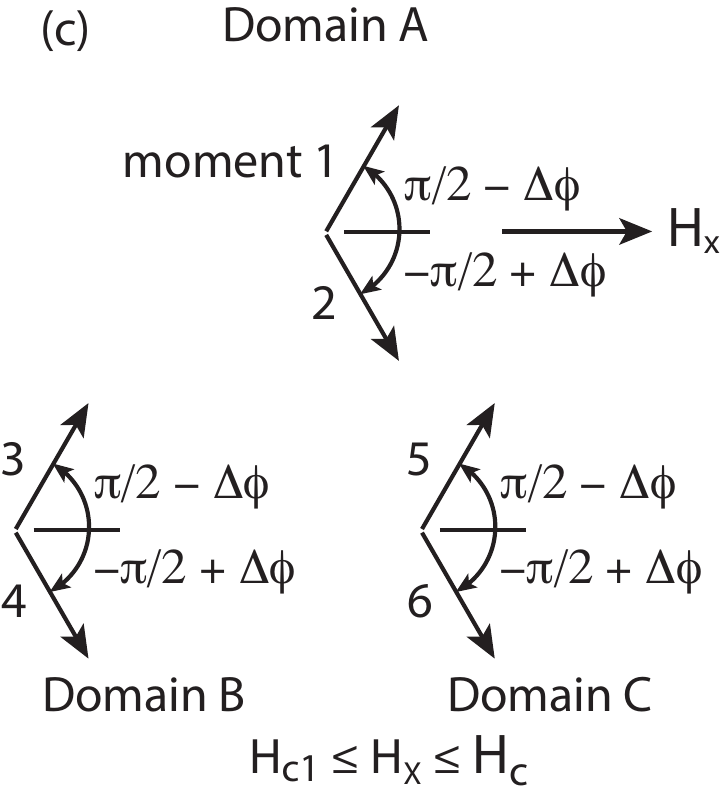}
\caption{(a) Domain model for trigonal A-type antiferromagnets in $H_x=0$~\cite{Pakhira2022b}.  Moments 1, 4, and 6 are the magnetic moments in a central layer in zero field according to Fig.~\ref{FigTrigAnisEnergy}, whereas moments 2, 3, and 5 represent moments in adjacent layers in the collinear A-type AFM structure.  The arrows and $\Delta\phi$ indicate the direction of rotation of the moments in a domain for $H_x\leq H_{{\rm c}1}$.  (b) Orientation of the moments when $H = H_{\rm c1}$, apart from a slight canting toward $H_x$ that gives rise to the measured magnetization.  (c)~Rotation of the moments in each domain towards $H_x$ for $H_{\rm c1} < H_x < H^{\rm c}_{ab}$, where $H^{\rm c}_{ab}$ is the $ab$-plane saturation (critical) field.}
\label{Fig_Domains}
\end{figure}

The related topological compounds \ems\ and \emb\ crystallize in the trigonal \cas-type crystal structure with space group $P\bar{3}m1$ and order antiferromagnetically at $T_{\rm N} = 8.2$ and~6.7~K, respectively.  Neutron-diffraction measurements demonstrated that both compounds exhibit A-type AFM order with the Eu moments aligned in the $ab$~plane~\cite{Pakhira2022, Pakhira2021a}.  Remarkably, crystals of both compounds showed a distinct positive curvature in $M(H_{ab})$ isotherms at low fields $0 \leq H_{ab}\lesssim H_{\rm c1}$ at $T=1.8$~K, where \mbox{$H_{\rm c1}\approx 220$} and $\approx 465$~Oe  for \ems\ and and \emb, respectively~\cite{Pakhira2022b}.   At higher fields up to the respective saturation critical fields which were \mbox{$H^{\rm c}_{c} = 3.4$~T} for \ems\ at 1.8~K~\cite{Pakhira2022} and 4.0~T at 2~K for \emb~\cite{Pakhira2020}, it was found that  $M_{ab}=\chi(T_{\rm N})H_{ab}$ as expected from molecular-field theory~\cite{Johnston2015} since the moments were initially nearly perpendicular to ${\bf H}_{ab}$ at the respective critical field $H_{\rm c1}$.

These observations motivated us to consider the presence of equally-populated trigonal AFM domains below the respective $T_{\rm N}$ of \emb\ and \ems.  These domains result from the existence of a trigonal anisotropy free energy of the form
\be
F(\phi) = K_3 \cos(3\phi),
\label{Eq:FofPhi}
\ee
where $K_3$ is the anisotropy energy amplitude and $\phi$ is the angle between moments in an A-type AFM domain and the $x$~axis, which is assumed to be the axis in which a magnetic field $H_x$ is applied.  A plot of $F(\phi)$ is shown in Fig.~\ref{FigTrigAnisEnergy}, where three  minima in $F(\phi)$ are at $\phi/\pi =  -5/6$, $-1/6$, and $1/2$.

The critical field $H_x=H_{\rm c1}$ is the $ab$-plane field at which all moments in each domains become nearly perpendicular to $H_{ab}$, apart from a small canting $\lesssim 1^\circ$ toward the field that gives rise to the observed magnetization.  The $M(0\leq H_{ab}\leq H_{\rm c1})$ data for both \ems\ and \emb\ were successfully fitted by a model in which the positive curvature in $M(H_{ab})$ below $H_{\rm c1}$ arose from the field response of the Eu spins in trigonal domains in Fig.~\ref{Fig_Domains} \cite{Pakhira2022b}.  These fits allowed the respective $ab$-plane trigonal anisotropy parameters $K_3$ to be determined.

\section{Theory}

\begin{figure}[ht]
\includegraphics[width = 2.5in]{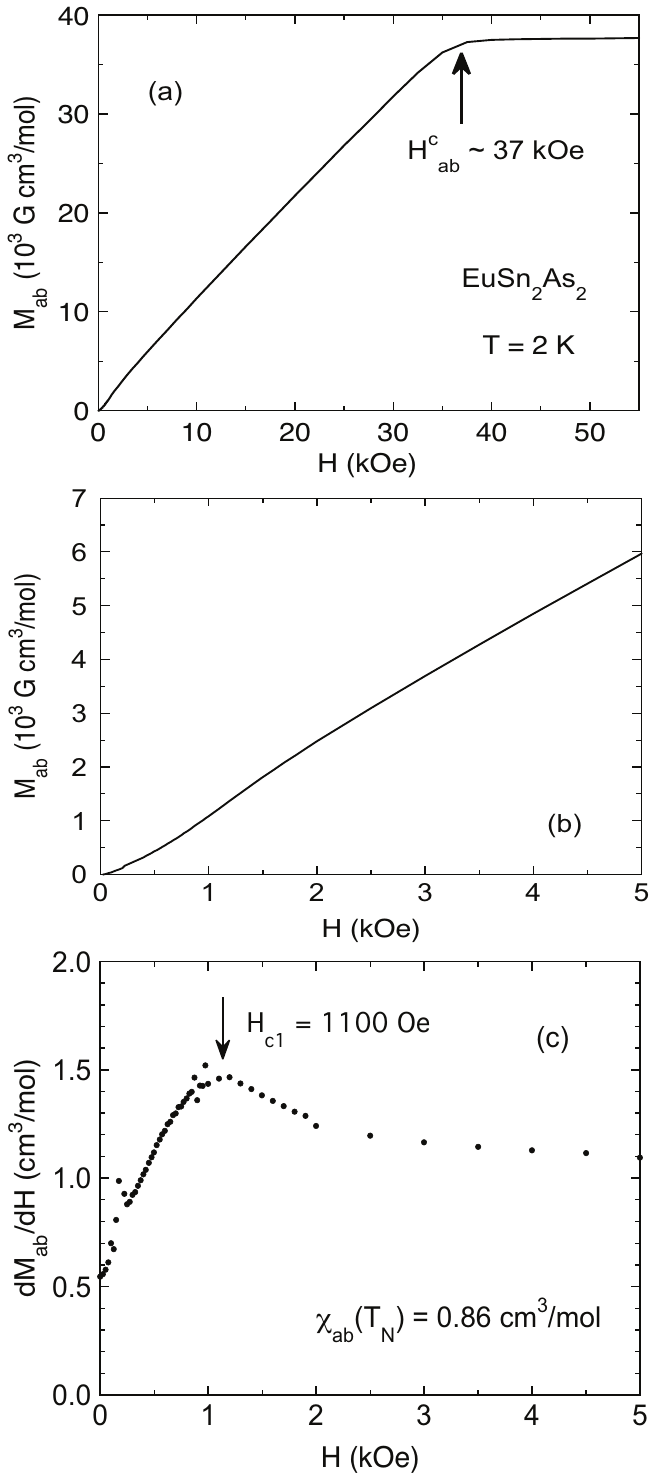}
\caption{(a) Magnetization versus $x$-axis field $H_x$ up to 55~kOe. The critical field for the $ab$-plane field is $\approx  37$~kOe.  (b) Expansion of $M_{ab}(H)$ for $H\leq 5$~kOe.  (c)~Low-field slope $dM_{ab}/dH$, yielding $H_{\rm c1}\approx 1100$~Oe. The small  peak at $H\approx 200$~Oe was also observed to occur for $dM_{ab}/dH$ with ${\bf H}\parallel c$~\cite{Pakhira2021} and was postulated to arise from the presence of magnetic defects.}
\label{Fig04}
\end{figure}

Here we analyze the low-field $M(H_{ab},T=2~{\rm K})$ data for \esa~\cite{Pakhira2021} in Fig.~\ref{Fig04}(b) by our model~\cite{Pakhira2022b}  for the magnetic response of the moments in isolated trigonal domains at $T = 0$~K and derive therefrom the value of its trigonal anisotropy constant $K_3$.  We will then compare this value with the values of $K_3$ previously obtained for trigonal domains in \ems\ and \emb\ \cite{Pakhira2022b}.

With increasing $ab$-plane field $H_x$ at $T=2$~K, the three AFM domains initially at 120$^\circ$ to each other as shown in Fig.~\ref{Fig_Domains}(a) begin to align nearly perpendicular to $H_x$, which according to Fig.~\ref{Fig04}(c) occurs  at $H_x\approx 1100$~Oe.  Then with a further increase in $H_x$ the collinear spins in each domain progressively cant towards $H_x=H_{ab}$ until they reach FM alignment at the critical field $H_{ab}^{\rm c} \approx 3.7$~T according to Fig.~\ref{Fig04}(a).

Following Ref.~\cite{Pakhira2022b}, the anisotropy energy (\ref{Eq:FofPhi}) averaged over the three domains and then normalized by $K_3$ is
\be
\frac{E_{\rm anis~ave}}{K_3} = -\frac{1}{3}[1+2\cos(3\Delta\phi)],
\label{Eq:EanisAve}
\ee
and the magnetic energy normalized by $K_3$ is
\be
\frac{E_{\rm mag\,ave}}{K_3} = -\frac{\chi_\perp H_x^2}{3K_3}\Big[1+2\sin^2\Big(\frac{\pi}{6}+\Delta\phi\Big)\Big],
\label{Eq:EmagAve}
\ee
where $\Delta\phi$ in Fig.~\ref{Fig_Domains}(c) is the rotation angle of the moments due to a particular $H_x$ and $\chi_\perp=\chi_c$ is the magnetic susceptibility perpendicular to the $ab$~plane, which according to molecular-field theory~\cite{Johnston2015} is $\chi_c = \chi(T_{\rm N})$.  Then setting 
\be
h_x=\chi_\perp H_x^2/K_3
\label{Eq:hxDef}
\ee
gives
\be
\frac{E_{\rm mag\,ave\,2}}{K_3} = -\frac{h_x}{3}\Big[1+2\sin^2\Big(\frac{\pi}{6}+\Delta\phi\Big)\Big].
\label{Eq:EmagAveOnK3}
\ee

Thus the total average energy $E_{\rm tot\,ave}$ normalized by $K_3$ is
\be
\frac{E_{\rm tot\,ave}}{K_3} = \frac{E_{\rm anis~ave}}{K_3} +\frac{E_{\rm mag\,ave\,2}}{K_3}.
\label{Eq:EmagTotAve}
\ee
Then using Eqs.~(\ref{Eq:EanisAve}) and~(\ref{Eq:EmagAveOnK3}) and minimizing $E_{\rm tot\,ave}/K_3$ in Eq.~(\ref{Eq:EmagTotAve}) with respect to $\Delta\phi$ gives a simple solution for $h_x$ in terms of $\Delta\phi$ as
\be
h_x(\phi) = 3\csc[(\pi+6\Delta\phi)/3]\sin(3\Delta\phi).
\label{Eq:HxDeltaphi}
\ee
From Figs.~\ref{Fig_Domains}(a) and \ref{Fig_Domains}(b), we have $\Delta\phi=\pi/3$ when $H_x=H_{\rm c1}$, {\it i.e.}, when all moments are approximately perpendicular to $H_x$.  Then Eqs.~(\ref{Eq:hxDef}) and~(\ref{Eq:HxDeltaphi})  and  give
\be
h_x(\pi/3) = \frac{9}{2} = \frac{\chi_\perp H_{\rm c1}^2}{K_3}.
\label{Eq:hx1}
\ee
This value is per Eu$^{2+}$ ion, whereas the measured \mbox{$\chi_\perp=\chi(T_{\rm N})$} is per mole of Eu ions, so Eq.~(\ref{Eq:hx1}) becomes 
\be
\frac{9}{2} = \frac{\chi_\perp H_{\rm c1}^2}{N_{\rm A}K_3},
\ee
where $N_{\rm A}$ is Avogadro's number.  Then using \mbox{$H_{\rm c1} = 1100$~Oe} from Fig.~\ref{Fig04}(c), the anisotropy-corrected $\chi_\perp= \chi(T_{\rm N})=0.86$~cm$^3$/mol~\cite{Pakhira2021}, and solving for $K_3$ gives
\bea
K_3 &=& \frac{\chi_\perp H_{\rm c1}^2}{(9/2)N_{\rm A}} \\
               &=& 2.4\times 10^{-7}~{\rm eV/Eu}.\nonumber
\eea
This $K_3$ value is 13.3 and 3.7 times larger than those found for single crystals of the two previously-studied trigonal A-type antiferromagnets \ems\ and \emb, respectively~\cite{Pakhira2022b}.

\section{Concluding Remarks}

The present work is part of an ongoing effort to characterize the $ab$-plane magnetic-field dependence of Eu$^{2+}$ and Gd$^{3+}$ spin $S=7/2$ trigonal and tetragonal \mbox{A-type} antiferromagnets with the moments aligned in the $ab$~plane.  In addition to the present study of \esa, these compounds have so far included trigonal \ems\ and \emb~\cite{Pakhira2022b} as noted in the text, and also trigonal EuAl$_2$Ge$_2$~\cite{EuAl2Ge2} and tetragonal EuGa$_4$~\cite{EuGa4}.    In future work, it would be interesting to determine if the $ab$-plane magnetic-field-induced changes in the domain structure also affects the topological properties of relevant A-type antiferromagnets.

\acknowledgments

The research at Ames National Laboratory was supported by the U.S. Department of Energy, Office of Basic Energy Sciences, Division of Materials Sciences and Engineering. Ames National Laboratory is operated for the U.S. Department of Energy by Iowa State University under Contract No.~DE-AC02-07CH11358.

\end{document}